\documentclass[10pt,twocolumn,letterpaper]{article}

\usepackage{cvpr}
\usepackage{times}
\usepackage{graphicx}
\usepackage{amsmath}
\usepackage{amssymb}
\usepackage{multirow}
\usepackage{subfig}
\usepackage{balance}

\usepackage[pagebackref=true,breaklinks=true,letterpaper=true,colorlinks,bookmarks=false]{hyperref}

\cvprfinalcopy % *** Uncomment this line for the final submission

\def\cvprPaperID{4} % *** Enter the CVPR Paper ID here

\ifcvprfinal\fi

\newcommand{\hide}[1]{}

\newcommand*\samethanks[1][\value{footnote}]{\footnotemark[#1]}
\begin{document}

%%%%%%%%% TITLE
\title{Localization of JPEG double compression\\
through multi-domain convolutional neural networks}

\author{Irene Amerini$^a$\thanks{indicates equal contribution and corresponding authors.}~, Tiberio Uricchio$^a$\samethanks~, Lamberto Ballan$^{a,b}$, Roberto Caldelli$^{a,c}$\medskip\\
$^a$Media Integration and Communication Center, University of Florence, Italy\\
$^b$Department of Mathematics ``Tullio Levi-Civita'', University of Padova, Italy\\
$^c$National Inter-University Consortium for Telecommunications (CNIT), Italy\\
%{\tt\small \{name.surname\}@unifi.it}
%\and
%Roberto Caldelli\\
%National Interuniversity Consortium for
%Telecommunications (CNIT), Parma, Italy\\
%Media Integration and Communication Center, University of Florence, Florence, Italy\\
%{\tt\small roberto.caldelli@unifi.it}
}

\maketitle
\thispagestyle{empty}

%%%%%%%%% ABSTRACT
\begin{abstract}
When an attacker wants to falsify an image, in most of cases she/he will perform a JPEG recompression. Different techniques have been developed based on diverse theoretical assumptions but very effective solutions have not been developed yet. Recently, machine learning based approaches have been started to appear in the field of image forensics to solve diverse tasks such as acquisition source identification and forgery detection. In this last case, the aim ahead would be to get a trained neural network able, given a to-be-checked image, to reliably localize the forged areas. With this in mind, our paper proposes a step forward in this direction by analyzing how a single or double JPEG compression can be revealed and localized using convolutional neural networks (CNNs). Different kinds of input to the CNN have been taken into consideration, and various experiments have been carried out trying also to evidence potential issues to be further investigated.
\end{abstract}

%%%%%%%%% BODY TEXT
\section{Introduction}

Nowadays the pervasiveness of images and also videos as primary source of information has led the image forensics community to question about their reliability and integrity more and more often. The context in which pictures are involved is disparate. A magazine, a social network, an insurance practice, an evidence for a trial. Such images can be easily altered through the use of powerful editing software, often leaving no visual trace of any modification, so answering reliably about their integrity becomes fundamental.
Image forensics deals with these issues by developing technological instruments which allow to determine, only on the basis of a picture, if that asset has been modified and sometimes to understand what has happened localizing the tampering. Regarding forgeries individuation three are the principal classes of detectors studied so far: those based on features descriptors \cite{ameriniIC, reiss2013, cozzolino2015splicebuster}, those based on inconsistent shadows \cite{Kee2013} and finally those based on double JPEG compression \cite{Lin-FastAutomatic,bianchi-Transaction,binli2008,milani2012,becarelliwifs2014}.

In recent years, machine learning and neural networks, such as convolutional neural networks (CNNs), have shown to be capable of extracting complex statistical features and to efficiently learn their representations, allowing to generalize well across a wide variety of computer vision tasks, including image recognition and classification and so on \cite{krizhevsky2012imagenet,girshick2014rich,simonyan2014very,jj2015iccv,uricchio2016automatic}.
The extensive use of such networks in many areas has motivated and led the multimedia forensics community to comprehend if such technological solutions can be employed to exploit source identification \cite{chaumont2016,BaroffioBBT16} or for image and video manipulation detection \cite{rao2016, rota2016, Bayar:2016, wang2016}. In particular, Wang \etal \cite{wang2016} use the histogram of Discrete Cosine Transform (DCT) coefficients as input to a CNN to detect single or double JPEG compressions in order to detect tampered images. The main idea behind \cite{rao2016, Bayar:2016} is to develop a sort of pre-processing module, designed to suppress image content before training a CNN; while, in \cite{rota2016} the CNN architecture is fed with patches without pre-processing and tampered patches are extracted from the borders of the tampered areas. Although the interest in  neural network in image forensics domain is growing, a real comprehension of what is possible to accomplish with it is still in an early stage.

\begin{figure*}[!t]
	\centering
	\includegraphics[width=1.\textwidth]{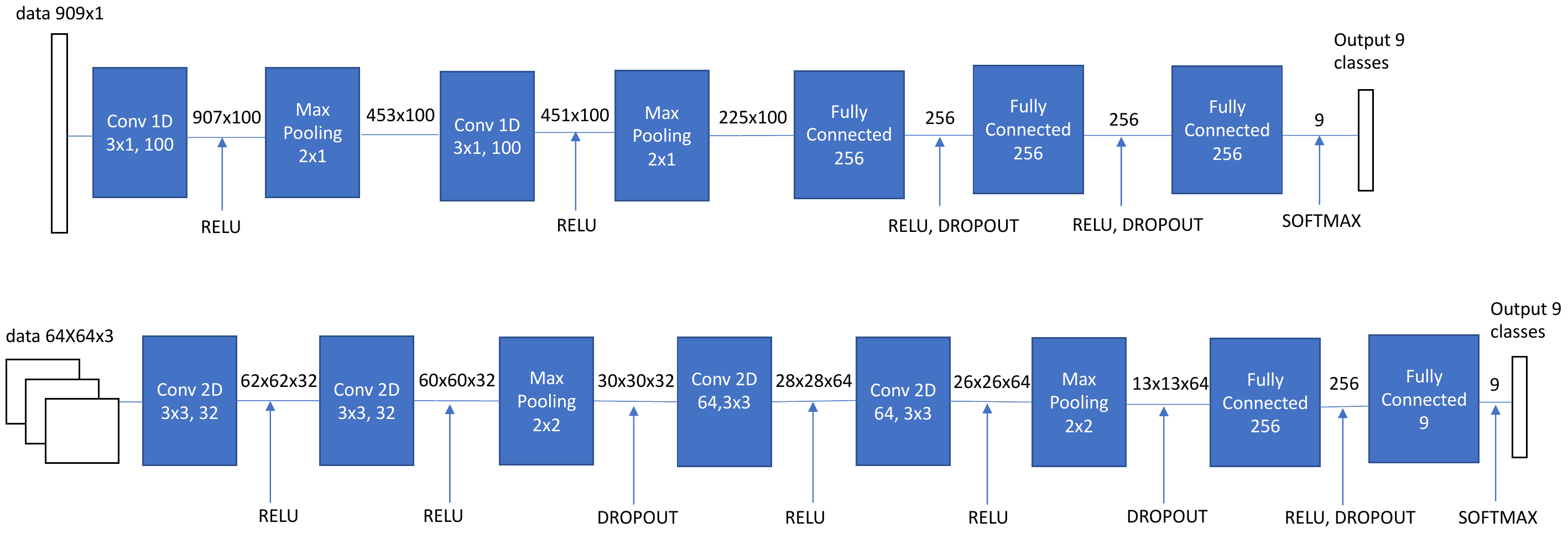}
	\caption{Architecture of the CNNs: \emph{frequency domain-based} CNN (top) and \emph{spatial domain-based} CNN (bottom).}
	\label{fig:CNN}
\end{figure*}

This paper presents a step forward in this direction. Our objective is to train a neural network that, given a to-be-checked image, is able to reliably localize the possible forged areas by analyzing the presence of single or double JPEG compressed areas.
In particular, different kinds of CNNs-based approaches have been proposed and different inputs to the nets are given. First of all, a \textit{spatial domain-based CNN} is exploited performing image forgery detection starting from the RGB color images; neither pre-processing is carried out nor side information on the borders of the tampered area is adopted. The CNN is trained to distinguish among uncompressed, single and double JPEG compressed images, to reveal the primary (hidden) JPEG compression and then localize the forgery regions. Secondly, another \textit{frequency domain-based CNN} is introduced taking as input to the net the histogram of the DCT coefficient similarly as \cite{wang2016}. The third proposed approach is a \textit{multi-domain-based CNN} able to join the two previous input information on RGBs patches and on DCT histograms.
The main contribution of this work is to explore the use of a spatial domain CNN and its combination with the frequency domain for the image forgery detection task. % of distinguish between the various JPEG compressions.
Disparate experimental tests have been carried out trying also to evidence potential issues to be further investigated and improved. 

The rest of the paper is organized as follows. In Section \ref{sec:approaches} we discuss the proposed approaches; Section \ref{sec:expres} contains experimental results, while conclusions and open issues are finally drawn in Section \ref{sec:conc}.

\section{CNN-based proposed approaches}
\label{sec:approaches}

In this work, our objective is to investigate the possibility to discern among uncompressed, single or double compressed images with the intent to detect image regions involved in a splicing attack. In addition to this, our secondary goal is to reveal the primary quality factor applied to the image or to the patch before the secondary compression is applied.
To accomplish this task three different CNN-based approaches are devised on the basis of the input data given to the net and on the net itself.
A convolutional neural network consists of several cascaded of convolutional layers and pooling layers followed by one or more fully-connected layers. Each considered CNN in the proposed approaches differs from the others in how components of the nets are combined together and from the number of layers employed, as described in detail in the following.
In order to learn discriminant features directly from data a consistent set of labeled images is needed in the training phase. For this reason, for all the considered approaches, images of different sizes are subdivided in patches (not overlapping) and then each of them is fed to the net.
Differently from the input, that it is different among the approaches, the output of the nets is the same. In particular, the three different proposed CNNs are able to discern among 9 classes: uncompressed, single compressed and double compressed patches (7 quality factors from 60 to 95, step by 5 is considered).

\subsection{Spatial-domain CNN}
In the first CNN-based approach, named \textit{spatial domain-based CNN}, the input of the net is a NxN size patches on the three color channels (RGB), pre-processing is not considered at all and only a normalization of the data (between 0 and 1) is performed.
First of all a convolutional network \cite{lecun1995convolutional} is designed and it is summarized in Figure \ref{fig:CNN} (top).
This particular net is composed by two convolutional blocks and two fully connected layers.
Each convolutional block is composed by two convolutional layers with ReLU activation followed by a pooling layer. All convolutional layers use a kernel size of 3x3 while pooling layer kernel size is 2x2. In order to prevent overfitting, we use Dropout \cite{srivastava2014dropout} that randomly drops units at training time from the fully connected layers.
In particular, a CNN of this kind is trained for each of the considered secondary quality factor $QF_2={60:5:95}$. Thus, we obtained eight different classifiers corresponding to each value of $QF_2$. 
Each classifier is required to output two levels of classifications for an input patch. The first is an inter-class categorization between uncompressed, single compressed and double compressed patch. The second is the intra-class of the $QF_1$ (ranging in ${60:5:95}$, excluding $QF_1=QF_2$) in the case of double compressed patches. We thus choose to output 9 plain classes, the uncompressed class, the single compressed class and a class for each $QF_1$.  
%The output for each classifier is a distribution of nine classes corresponding to the uncompressed, single compressed and double compressed patches (with $QF_2$ fixed and primary quality factors $QF_1$ varying in ${60:5:95}$, $QF_1=QF_2$ falls in the case ``single"). 
As a result, the last fully connected layers of the CNN is sent to a nine-way softmax connection, which produces the probability that each sample should be classified into each class. As loss function, we use a categorical cross-entropy function \cite{vincent2010stacked}. We note that mis-classifying the intra-class of a double compressed patch is a smaller error compared to wrongly classify the inter-class of a patch. So, we adjust the loss to weight an intra-class error as $1/9$ of an inter-class error. In our preliminary experiments, this improved the intra-class classification accuracy.

The proposed CNN model is trained based on the labeled patch samples from the training set composed by uncompressed, single or double compressed patches (i.e $QF_2=90$ and $QF_1$ varies from 60 to 95). In the test phase, one of the eight trained CNN (selected accordingly to the quality factor saved in EXIF header of the JPEG format) is used to extract the patch-based features for a test image by applying a patch-sized sliding window to scan the whole image, assigning a class for each patch and therefore performing localization at image level.

\subsection{Frequency-domain CNN}
In the second proposed approach, \textit{frequency domain-based CNN}, a pre-processing is performed for a given patch computing the histogram of the DCT coefficients following the idea in \cite{wang2016} expanding the number of the evaluated coefficients.
In detail, given a NxN patch, DCT coefficients are extracted and, for each 8x8 block, the first 9 spatial frequencies in zig-zag scan order (DC is skipped) are selected. For each spatial frequency $i,j$, the histogram $h_{i,j}$, representing the occurrences of absolute values of quantized DCT values, is built. In detail, $h_{i,j}(m)$ is the number of values $m$ in the histogram of the $i,j$ DCT coefficient with $m=(-50..,0..,+50)$.
So the network take in total a vector of 909 elements (101 histogram bins x 9 DCT frequencies) as input.
Again, as before, an array of eight CNNs is trained, each of them corresponding at the different values of the second compression quality factor $QF_2$.
The feature vector is then used to train each CNN, in order to distinguish among the 9 classes defined before (uncompressed, single compressed and double compressed with $QF_2$ fixed and primary quality factors varying in $QF_1={60:5:95}$).
The architecture of the proposed CNN model is illustrated in Figure \ref{fig:CNN} (bottom). It contains two convolutional layers followed by two pooling connections and three full connections. The size of the input data is 909x1, and the output is a distribution of nine classes.
Each fully connected layer has 256 neurons, and the output of the last one is sent to a nine-way softmax, which produces the probability that each sample should be classified into each class. In our network, rectified linear units (ReLUs) $f(x) = max(0, x)$ as activation function, are used in each layer. In both fully
connected layers, the Dropout technique is used.

\subsection{Multi-domain CNN}
The third considered approach is a \textit{multi-domain CNN} where the three channels color patch and the histogram of DCT coefficient computed on the patch serve as input of the net in order to combine the previous two approaches.
\begin{figure}[!t]
	\centering
	\includegraphics[width=0.9\columnwidth]{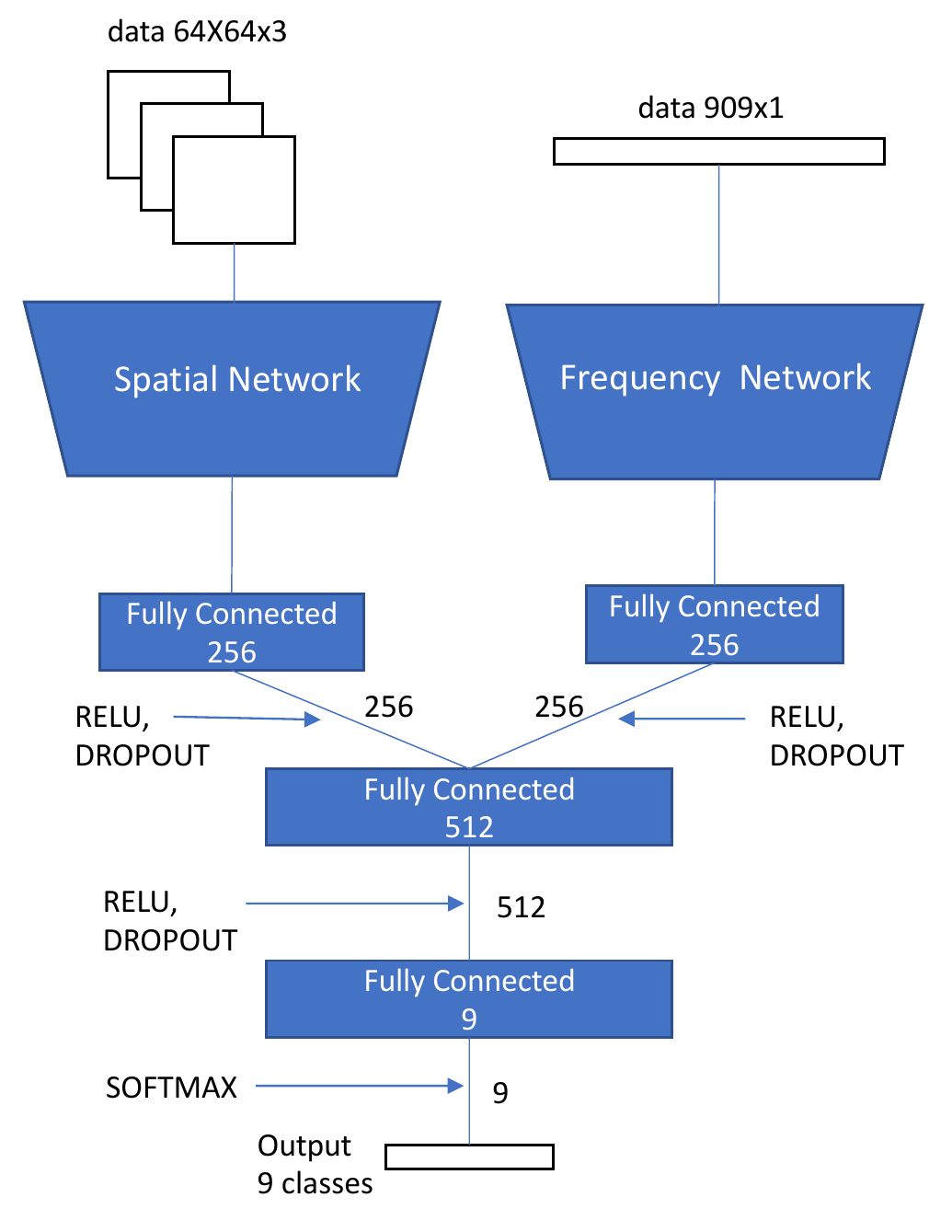}
	\caption{Architecture of the \emph{multi-domain} CNN.}
	\label{fig:mCNN}
\end{figure}
In Figure \ref{fig:mCNN} the proposed net is depicted and it consists of one \textit{spatial domain-based CNN} and one \textit{frequency domain-based CNN} up to their first respective fully connected layers. The \textit{multi-domain-based CNN} learns the inter-modal relations between features coming from R,G,B domain and from the histogram of DCT joining together the outputs of the fully connected layers of the two nets (256 dimensions each). In this way the last fully connected layer has 512 neurons, and the output is sent to a nine-way softmax connection, which produces the probability that each sample is classified into each class also using a dropout layer.
So, as well as before, eight different 9 classes classifiers are devised corresponding to each value of $QF_2$. The training and testing phase are performed as before.

\section{Experiments}
\label{sec:expres}
In this section some of the experimental tests carried out are presented. In particular, in Section \ref{ssec:setup}
the general set-up is primarily introduced while in Section \ref{ssec:exp_9class} results obtained with the 9-classes classifiers are presented and in Section \ref{ssec:exp_9class}, the performance of the three proposed approaches are compared. Ultimately, in Section \ref{ssec:exp_quality}, a qualitative point of view on some forensic-like examples is debated.

\subsection{Experimental setup}
\label{ssec:setup}
The UCID dataset \cite{Schaefer04ucid} has been used for the experimental tests; it is composed by 1338 images (TIFF format and size $384\times512$). The whole dataset has been subdivided in training set (1204, about $90\%$), validation set (67, about $5\%$) and test set (67, about $5\%$) in order to keep separate the bunches of images involved in the different phases.
It have been considered 8 diverse JPEG quality factors with $QF=60:95$ with a step of $5$ both for the first and the second compression; according to this, 8 CNNs (one for each $QF_2$) have been trained on non-overlapping image patches of size $N=64$ that is $48$ patches for every UCID image. Each CNN is trained to classify 9 different classes of images which are: \emph{uncompressed}, \emph{single compressed} and \emph{double compressed} (7 classes, given that the case $QF_1=QF_2$ is skipped because it would fall in the single compressed one). The neural network learns on 57,792 patches ($1204\times48$) for each of the 9 classes and is optimized by using AdaDelta method \cite{zeiler2012adadelta}. The training phase is stopped when the loss function on the validation set reaches its minimum that usually happens after $15/20$ epochs. Performance on the test set (28,944 patches in total) are evaluated in terms of \emph{True Positive Rate} ($TPR=\frac{TP}{TP+FN}$)
%, as for the tables in Section \ref{ssec:exp_9class} % 
and \emph{Accuracy} ($ACC=\frac{TP+TN}{TP+FN+TN+FP}$). % like in Section \ref{ssec:exp_9class}.

\subsection{CNN-based approaches evaluation}
\label{ssec:exp_9class}
In this experiment, we have investigated the performance in terms of TPR of the CNNs trained with spatial domain-based examples and with frequency domain-based ones. Results, over a test-set of 28,944 image patches, of the different CNNs are presented in Table \ref{tab:CM1_image} and Table \ref{tab:CM1_DCT} respectively. 
Both methods are able to classify all uncompressed patches almost perfectly while the spatial domain-based CNN has an higher TPR for single compressed patches.  
Regarding double compressed patches, it can be seen that both methods show good performance in the top-right zone of the matrix. It is quite well-known in fact that when $QF_2 > QF_1$, traces of the first compression still survive and are easily detectable.

\begin{table*}[]
\centering
\begin{tabular}{|c|l|c|c|c|c|c|c|c|c|c|}
\hline
\multirow{2}{*}{} & & \multicolumn{8}{|c|}{$QF_2$} & \\ \cline{2-11}
&  & \textbf{60} & \textbf{65} & \textbf{70} & \textbf{75} & \textbf{80} & \textbf{85} & \textbf{90} & \textbf{95} & \textbf{AVG} \\ \hline \hline
\multirow{2}{*}{} & \textbf{Uncompressed}  & 0.999 & 0.998 & 0.994 & 0.997 & 0.997 & 0.999 & 0.997 & 0.996 & 0.997\\ \cline{2-11}
& \textbf{Single Compressed}  & 0.599 & 0.701 & 0.717 & 0.789 & 0.843 & 0.955 & 0.981 & 0.986 & 0.821\\ \hline
\multirow{8}{*}{\rotatebox[origin=c]{90}{$QF_1$}} & \textbf{60}  & --- & 0.403 & 0.870 & 0.918 & 0.804 & 0.912 & 0.803 & 0.827 & 0.791\\ \cline{2-11}
& \textbf{65}  & 0.235 & --- & 0.470 & 0.783 & 0.532 & 0.672 & 0.669 & 0.771 & 0.590\\ \cline{2-11}
& \textbf{70}  & 0.423 & 0.356 & --- & 0.555 & 0.646 & 0.551 & 0.661 & 0.819 & 0.573\\ \cline{2-11}
& \textbf{75}  & 0.633 & 0.561 & 0.415 & --- & 0.746 & 0.716 & 0.739 & 0.785 & 0.656\\ \cline{2-11}
& \textbf{80}  & 0.796 & 0.714 & 0.580 & 0.467 & --- & 0.891 & 0.810 & 0.852 & 0.730\\ \cline{2-11}
& \textbf{85}  & 0.636 & 0.469 & 0.792 & 0.826 & 0.794 & --- & 0.908 & 0.926 & 0.764\\ \cline{2-11}
& \textbf{90}  & 0.740 & 0.755 & 0.771 & 0.746 & 0.899 & 0.956 & --- & 0.991 & 0.837\\ \cline{2-11}
& \textbf{95}  & 0.702 & 0.713 & 0.395 & 0.734 & 0.896 & 0.932 & 0.942 & --- & 0.759\\ \hline
\end{tabular}
\caption{\emph{Spatial domain-based} CNNs: performance of the 8 CNNs to distinguish the 9 different classes of images in terms of TPR.}
\label{tab:CM1_image}
\end{table*}

\begin{table*}[]
\centering
\begin{tabular}{|c|l|c|c|c|c|c|c|c|c|c|}
\hline
\multirow{2}{*}{} & & \multicolumn{8}{|c|}{$QF_2$} & \\ \cline{2-11}
&  & \textbf{60} & \textbf{65} & \textbf{70} & \textbf{75} & \textbf{80} & \textbf{85} & \textbf{90} & \textbf{95} & \textbf{AVG} \\ \hline \hline
\multirow{2}{*}{} & \textbf{Uncompressed}  & 1.000 & 0.999 & 1.000 & 1.000 & 1.000 & 0.999 & 0.999 & 0.998 & 0.998\\ \cline{2-11}
& \textbf{Single Compressed}  & 0.490 & 0.395 & 0.472 & 0.717 & 0.668 & 0.765 & 0.874 & 0.995 & 0.672\\ \hline
\multirow{8}{*}{\rotatebox[origin=c]{90}{$QF_1$}} & \textbf{60}  & --- & 0.886 & 0.938 & 0.991 & 0.991 & 0.992 & 0.994 & 0.995 & 0.970\\ \cline{2-11}
& \textbf{65}  & 0.647 & --- & 0.868 & 0.944 & 0.959 & 0.972 & 0.972 & 0.979 & 0.906\\ \cline{2-11}
& \textbf{70}  & 0.876 & 0.571 & --- & 0.873 & 0.958 & 0.977 & 0.984 & 0.982 & 0.889\\ \cline{2-11}
& \textbf{75}  & 0.824 & 0.907 & 0.743 & --- & 0.970 & 0.976 & 0.982 & 0.987 & 0.913\\ \cline{2-11}
& \textbf{80}  & 0.727 & 0.765 & 0.910 & 0.894 & --- & 0.979 & 0.991 & 0.994 & 0.894\\ \cline{2-11}
& \textbf{85}  & 0.806 & 0.658 & 0.657 & 0.881 & 0.902 & --- & 0.984 & 0.986 & 0.839\\ \cline{2-11}
& \textbf{90}  & 0.450 & 0.388 & 0.574 & 0.723 & 0.802 & 0.913 & --- & 0.991 & 0.692\\ \cline{2-11}
& \textbf{95}  & 0.120 & 0.189 & 0.226 & 0.015 & 0.220 & 0.524 & 0.772 & --- & 0.295\\ \hline
\end{tabular}
\caption{\emph{Frequency domain-based} CNNs: performance of the 8 CNNs to distinguish the 9 different classes of images in terms of TPR.}
\label{tab:CM1_DCT}
\end{table*}

In Table \ref{tab:CM1_image_DCT}, the results obtained for the multi-domain-based approach which combines the two previous kinds of input, are listed. It is worthy underlining that there is a significant improvement, as general, and also in the bottom-left part of the table ($QF_1 < QF_2$). This suggests that the two inputs provide complementary information that the multi-domain approach is able to correlate and exploit. 

\begin{table*}[]
\centering
\begin{tabular}{|c|l|c|c|c|c|c|c|c|c|c|}
\hline
\multirow{2}{*}{} & & \multicolumn{8}{|c|}{$QF_2$} & \\ \cline{2-11}
&  & \textbf{60} & \textbf{65} & \textbf{70} & \textbf{75} & \textbf{80} & \textbf{85} & \textbf{90} & \textbf{95} & \textbf{AVG} \\ \hline \hline
\multirow{2}{*}{} & \textbf{Uncompressed}  & 0.999 & 1.000 & 1.000 & 0.999 & 0.999 & 1.000 & 1.000 & 0.997 & 0.999\\ \cline{2-11}
& \textbf{Single Compressed}  & 0.833 & 0.833 & 0.881 & 0.903 & 0.949 & 0.971 & 0.965 & 0.994 & 0.916\\ \hline
\multirow{8}{*}{\rotatebox[origin=c]{90}{$QF_1$}} & \textbf{60}  & --- & 0.892 & 0.983 & 0.987 & 0.976 & 0.990 & 0.980 & 0.992 & 0.972\\ \cline{2-11}
& \textbf{65}  & 0.679 & --- & 0.864 & 0.958 & 0.946 & 0.980 & 0.994 & 0.991 & 0.916\\ \cline{2-11}
& \textbf{70}  & 0.838 & 0.760 & --- & 0.869 & 0.985 & 0.982 & 0.981 & 0.985 & 0.914\\ \cline{2-11}
& \textbf{75}  & 0.864 & 0.809 & 0.734 & --- & 0.952 & 0.974 & 0.982 & 0.991 & 0.901\\ \cline{2-11}
& \textbf{80}  & 0.772 & 0.880 & 0.841 & 0.884 & --- & 0.982 & 0.994 & 0.991 & 0.907\\ \cline{2-11}
& \textbf{85}  & 0.764 & 0.760 & 0.815 & 0.876 & 0.919 & --- & 0.984 & 0.993 & 0.873\\ \cline{2-11}
& \textbf{90}  & 0.659 & 0.767 & 0.739 & 0.826 & 0.924 & 0.955 & --- & 0.996 & 0.838\\ \cline{2-11}
& \textbf{95}  & 0.830 & 0.703 & 0.773 & 0.760 & 0.932 & 0.959 & 0.889 & --- & 0.835\\ \hline
\end{tabular}
\caption{\emph{Multi-domain-based} CNNs: performance of the 8 CNNs to distinguish the 9 different classes of images in terms of TPR.}
\label{tab:CM1_image_DCT}
\end{table*}

%\subsection{3classes}
%\label{ssec:exp_3class}
%Uncompressed vs Single vs Double

\begin{figure}[!t]
	\centering
	\includegraphics[width=\columnwidth]{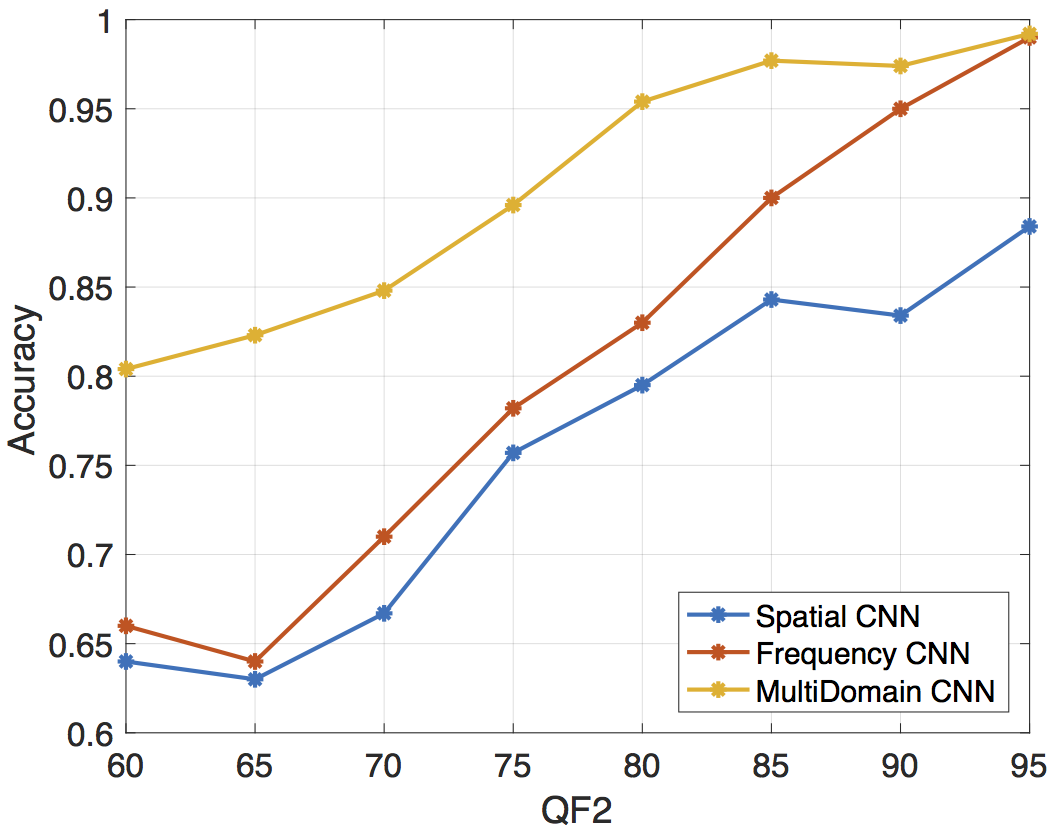}
	\caption{Three approaches comparison in terms of accuracy for each of the 8 ($QF_2$) classifiers.}
	\label{fig:comparison}
\end{figure}

%\subsection{Performance comparison}
%\label{ssec:exp_combo}
The three approaches are also compared in terms of accuracy for the different 8 classifiers according to $QF_2$. Figure \ref{fig:comparison} provides a clear evidence of the respective behaviors: the multi-domain approach outperforms the others and basically tends to achieve high level of accuracy (over $95\%$) when $QF_2$ is superior to the value of $80$. 
%The frequency-based CNN has higher accuracy than spatial-based CNN for every $QF_2$. Since the  suggesting that the network 

\subsection{Qualitative results}
\label{ssec:exp_quality}
\begin{figure*}[!htb]
\centering
%\vspace{1em}
\subfloat[]{%
\includegraphics[width=0.19\textwidth]{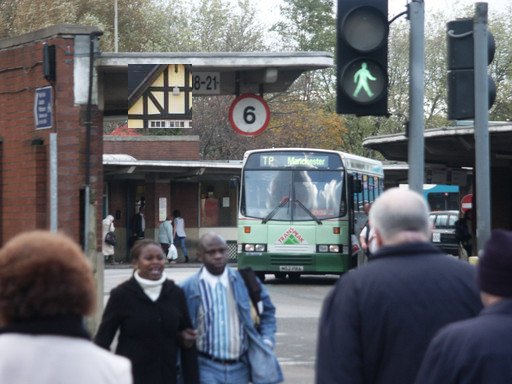}
}
\subfloat[]{%
 	\includegraphics[width=0.19\textwidth]{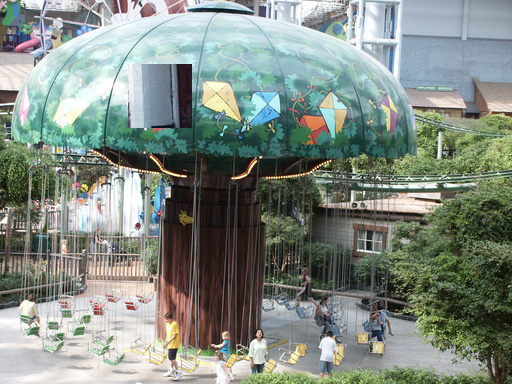}
}
\subfloat[]{%
 	\includegraphics[width=0.19\textwidth]{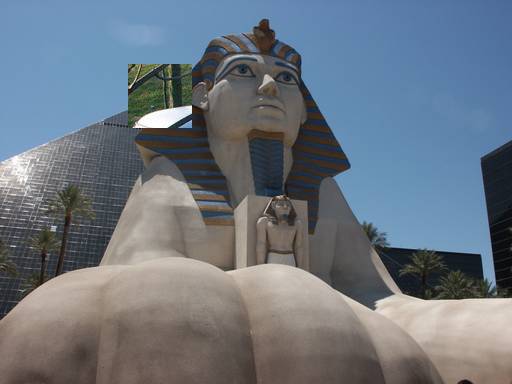}
}
\subfloat[]{%
 	\includegraphics[width=0.19\textwidth]{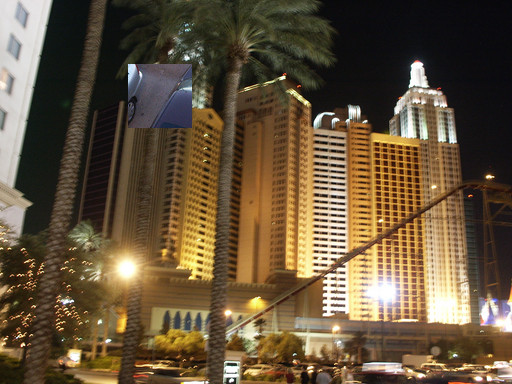}
}
\subfloat[]{%
 	\includegraphics[width=0.19\textwidth]{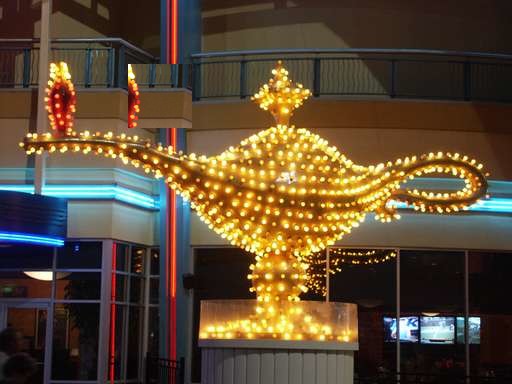}
}
\\
\subfloat[]{%
\includegraphics[width=0.19\textwidth]{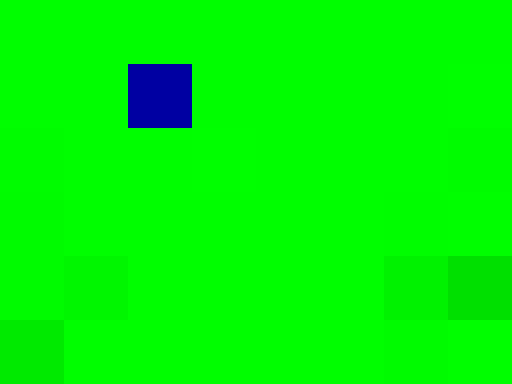}
}
\subfloat[]{%
 	\includegraphics[width=0.19\textwidth]{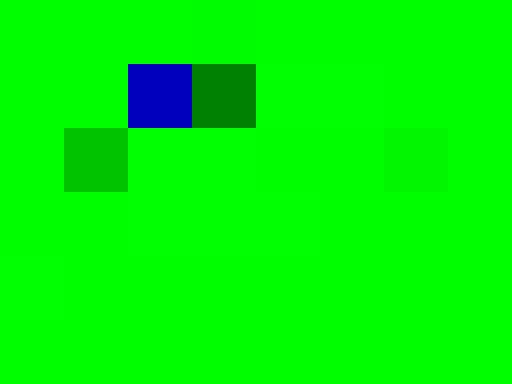}
}
\subfloat[]{%
 	\includegraphics[width=0.19\textwidth]{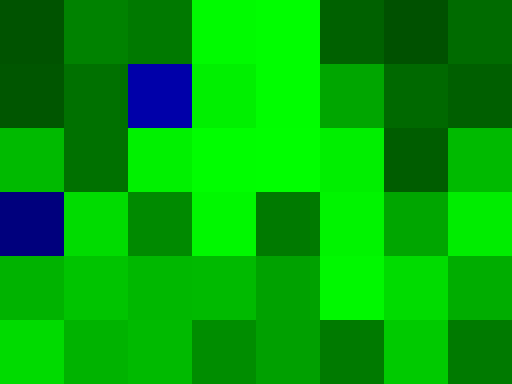}
}
\subfloat[]{%
 	\includegraphics[width=0.19\textwidth]{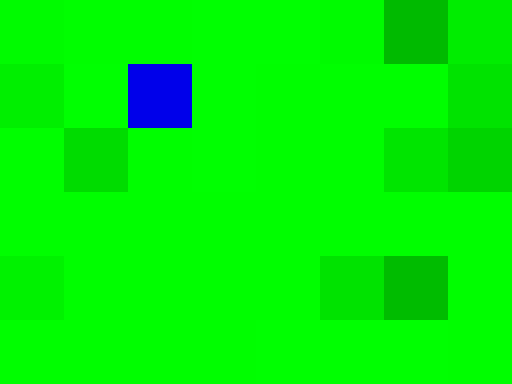}
}
\subfloat[]{%
 	\includegraphics[width=0.19\textwidth]{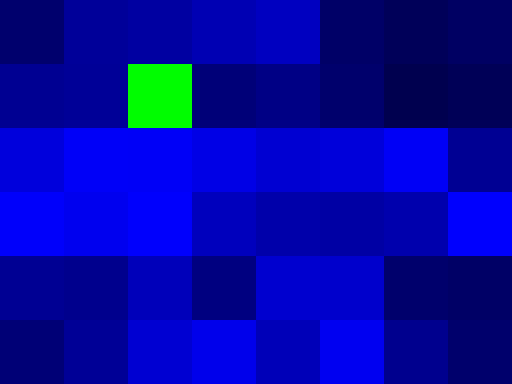}
}
\caption{Examples of forged images (top row) and corresponding localization masks (bottom row).}
\label{fig:masks}
\end{figure*}

In this section, some experimental results are extrapolated and presented to provide a qualitative view of the achieved performance mainly in terms of forgery localization. In particular, in Figure \ref{fig:masks} five sample counterfeited pictures (top row) and their corresponding localization masks (bottom row) are visualized. Forged images have been constructed by inserting a $64\times64$ patch, coming from another UCID image, within a host picture (for sake of clarity, the patch is located always in the same position in this figure). Such a processing can be carried out in different manners in terms both of used JPEG quality factors and of areas undergone to single or double compression; to provide an as-wide-as-possible view of the various cases diverse situations are represented. In Figure \ref{fig:masks} (a) and (f), the forged patch was double compressed (blue color) with $QF_1=60$ and $QF_2=90$ while the remaining part was single compressed (green color) at $QF_2=90$; different color tones indicate prediction probability of that class assigned by the CNN. So in this initial case, the second JPEG quality factor is higher than the first. In Figure \ref{fig:masks} (b) and (g), a similar case is considered but now $QF_1=80$ and $QF_2=85$, so quality factors are again in increasing order but much closer each other. On the contrary, in Figure \ref{fig:masks} (c)-(h) and (d)-(i), quality factors in decreasing order have been used ($QF_1=80$, $QF_2=70$ and $QF_1=95$, $QF_2=90$ respectively). It can be seen that now, as expected, the behavior is more noisy especially when the second compression is stronger ($QF_2=70$). Finally, in Figure \ref{fig:masks} (e) and (j), the case with $QF_1=60$ and $QF_2=90$ is presented but, this time, the forged patch is single ($QF_2=90$) compressed (green color). This is the dual circumstance, in terms of areas involved in compression, with respect to Figure \ref{fig:masks} (a) and (f).

\section{Conclusions}
\label{sec:conc}
In this paper we presented a step forward into adopting convolutional neural networks for the task of detecting splicing forgery. We began to explore CNN capabilities to classify and localize uncompressed, single and double compressed patches of images. In the latest case, our approach is also able to recover the original compression quality factor. We proposed a \textit{spatial domain-based CNN} and its combination with a \textit{frequency-based CNN} into a \textit{multi-domain-based} approach. Experimental results suggest that the spatial domain can be used directly and, when combined with the frequency domain, can lead to superior performance where DCT methods are usually weak (e.g. $QF_2 < QF_1$). 

Some open issues remain to be explored. First, the choice of the CNN architecture can lead to very different performance as it was seen on the object classification task \cite{krizhevsky2012imagenet, simonyan2014very} where deeper architectures are used.
Second, how much data is needed to train a good CNN model should be explored by collecting a larger dataset. Our results suggest that spatial information could help where DCT methods require patches with at least 64x64 to build a useful statistic.
Third, the capability of CNNs to detect different kind of compressions (e.g. JPEG 2000 or lossy PNG) should be explored. Our promising results show that this tool can detect the subtleties features of previous compressions and learn to predict the first quality factor used in re-compressions. 

{\small
\bibliographystyle{ieee}
\bibliography{forensic_ref}
}

\end{document}